\def\spose#1{\hbox to 0pt{#1\hss}}
\def\approxlt{\mathrel{\spose{\lower 3pt\hbox{$\sim$}}
	\raise 2.0pt\hbox{$$<$$}}}
\def\approxgt{\mathrel{\spose{\lower 3pt\hbox{$\sim$}}
	\raise 2.0pt\hbox{$>$}}}

\def\multleft#1{\hbox to size{\vbox {\halign {\lft{##}\cr #1}}\hfill}\par}
\def\multright#1{\hbox to size{\vbox {\halign {\rt{##}\cr #1}}\hfill}\par}

\def\today{\ifcase\month\or January\or February\or March\or April\or May\or
      June\or July\or August\or September\or October\or November\or December\fi
      \space\number\day, \number\year}
\def\$<${\thinspace}
\def\s{\hbox{\phantom{5}}}	%one space
\def\boxit#1{\vbox{\hrule\hbox{\vrule\kern3pt\vbox{\kern3pt
          #1 \kern3pt}\kern3pt\vrule}\hrule}}

\def\cm{{\rm\thinspace cm}}

\def\erg{{\rm\thinspace erg}}

\def\K{{\rm\thinspace K}}
\def\keV{{\rm\thinspace keV}}
\def\km{{\rm\thinspace km}}

\def\Mpc{{\rm\thinspace Mpc}}

\def\pc{{\rm\thinspace pc}}

\def\s{{\rm\thinspace s}}

\def\ergpcmsqps{\hbox{$\erg\cm^{-2}\s^{-1}\,$}}

\def\ergps{\hbox{$\erg\s^{-1}\,$}}

\def\kmps{\hbox{$\km\s^{-1}\,$}}

\def\psqcm{\hbox{$\cm^{-2}\,$}}

\def\kmpspMpc{\hbox{$\kmps\Mpc^{-1}$}}

\def\micron{{\rm\thinspace $\mu$m}}
\def\microJy{{\rm\thinspace $\mu$Jy}}

\documentstyle[psfig]{mn}
\begin{document}
\hsize=6truein

\title{Hot dust in two hard {\em Chandra} X-ray sources}

\author[]
{\parbox[]{6.in} {R.J.~Wilman, A.C.~Fabian and P. Gandhi\\
\footnotesize
Institute of Astronomy, Madingley Road, Cambridge CB3 0HA \\ }}

\maketitle

\begin{abstract}
The two brightest hard X-ray sources discovered serendipitously by {\em Chandra} in the field of the lensing cluster A2390 are found to have {\em ISOCAM} counterparts at 6.7 and 15\micron. We use this fact, together with their non-detection by {\em SCUBA} at 850\micron, as the basis for DUSTY radiative transfer modelling of their infrared spectral energy distributions. For the best-fitting models, we find that the dust which reprocesses the optical--UV light in these Compton-thin AGN is heated to near its sublimation temperature (above 1000\K), with an inner radius within a parsec of the nucleus. Some warm dust models with inner temperatures of 200\K~are also formally acceptable. These findings strongly support the obscured AGN hypothesis for the new hard X-ray {\em Chandra} sources, which lack both strong emission lines and 850\micron~{\em SCUBA} detections. 
\end{abstract}


\section{INTRODUCTION}
The {\em Chandra} X-ray observatory has recently resolved the bulk of the 2--7\keV~hard X-ray background (HXB) into point sources (Mushotzky et al.~2000; Brandt et al.~2000). Much effort is now focused on determining the physical nature of these new sources and their relation to existing classes of active galactic nuclei (AGN). Mushotzky et al.~(2000) found that roughly a third of their sources were blue broad-line (type I) AGN, one third were identified with apparently normal galaxies, whilst the final third had an extremely faint or no optical counterpart ($I \gg 23$mag) (see also Fiore et al.~2000). Crawford et al.~(2000) confirmed that sources in the last group are easily detected in the near-infrared, with featureless spectra suggesting that the AGN is heavily obscured and colours consistent with reddened elliptical galaxies at $z=1-2$. Further insight has come from cross-correlating {\em Chandra} source lists with those from 850\micron~{\em SCUBA} observations of the same fields: for the lensing clusters A2390 and A1835, Fabian et al.~(2000) (hereafter F00) found only one source common to both datasets, implying that any AGN in the {\em SCUBA} sources must either contribute little to the sub-mm power or be Compton-thick with X-ray scattering fractions below 1 per cent; this result was confirmed by Hornschemeier et al.~(2000), who detected none of the 10 sub-mm sources in and around the Hubble Deep Field North (HDFN) in a 166-ks {\em Chandra} exposure.

The two brightest X-ray sources in the A2390 field, CXOUJ215334.0+174240 and CXOUJ215333.2+174209 (hereafter sources A and B respectively), have {\em ISOCAM} counterparts at 6.7 and 15\micron~in the literature (Altieri et al.~1999; L\'{e}monon et al.~1998). Together with their non-detection at 850\micron~by {\em SCUBA}, this has implications for the location and properties of the dust which may be associated with the X-ray obscuration. In this paper we use the dust radiative transfer code DUSTY (Ivezi\'{c}, Nenkova \& Elitzur~1999) to model the infrared spectral energy distributions (SEDs) of these two sources as reprocessed optical--UV emission from an AGN. The faint optical light from these sources ($V\sim 25$) is assumed to be from the host galaxy and the HYPERZ code (Bolzonella, Pell\'{o} \& Miralles~2000) is used to estimate photometric redshifts.

\section{SED modelling}
For the primary optical--UV AGN continuum input to the DUSTY models, we follow Granato, Danese \& Franceschini~(1996) and use a broken power-law with $\alpha=-0.5$ for $\rm{log~\nu} < 15.4$, $\alpha=-1.0$ for $15.4 \leq \rm{log~\nu} < 16$, and $\alpha=-2.2$ for $\rm{log~\nu} \geq 16$ ($L_{\rm{\nu}} \propto \nu^{\alpha}$). The central source is surrounded by a spherical dust cloud, specified by its optical depth $\tau$ at 0.3\micron~($A_{\rm{V}} \simeq 0.57\tau$, and for a Galactic dust:gas ratio $N_{\rm{H}} \simeq 1.1 \times 10^{21}\tau$\psqcm), the ratio $R=r_{\rm{out}}/r_{\rm{in}}$ of the radii of the outer and inner edges of the distribution (between which we take the density to be constant), and the dust temperature $T_{\rm{in}}$ at $r_{\rm{in}}$. The code currently only supports grains of a single type, so we adopt the default `standard ISM mixture' comprising grains which mimic a 53:47 mixture of silicate and graphite grains from Draine \& Lee~(1984). The DUSTY wavelength grid spans 0.01\micron--$3.6 \times 10^{4}$\micron, so the effects of harder radiation are not modelled. Since sources A and B are likely to be Compton-thin with $N_{\rm{H}} \sim 10^{22}-10^{23}$\psqcm~(as inferred from the hardness ratios in F00 and from AGN synthesis models for the XRB, e.g. Wilman \& Fabian~1999), they are transparent to photons above a few keV; there are thus about 1.5 decades of frequency below 0.01\micron~where photons would be absorbed but which are not included. Given that the input SED cuts off sharply below 0.03 \micron, this omission is unimportant. 

We adopt the cosmological parameters $H_{\rm{0}}=50$\kmpspMpc~and $q_{\rm{0}}=0.5$ throughout the paper. 

\subsection{Source B (CXOUJ215333.2+174209)}
L\'{e}monon et al.~(1998) give 6.7 and 15\micron~{\em ISOCAM} fluxes for source B of $110^{+40}_{-60}$ and $350^{+50}_{-40}$\microJy, respectively. From a total of 33 counts, F00 estimate 0.5--2~and 2--7\keV~fluxes of 5.9 and 23$\times 10^{-15}$\ergpcmsqps, respectively; the upper limit at 850\micron~is 5.7~mJy (F00). We use the relative fluxes in the B, V(F555W), R, I(F814W), J and K$'$ filters provided by L\'{e}monon et al. along with HYPERZ to compute a photometric redshift (N.B. the HST F555W and F814W magnitudes given by L\'{e}monon et al. differ from those listed in F00, but for internal consistency we adopt the former values). We consider 2 families of HYPERZ models: (i) where at least 80 per cent of the K$'$ light (2.103\micron) is host galaxy starlight; (ii) where at least 50 per cent of it is nuclear dust emission. A satisfactory HYPERZ model could not be found for the case where all the K$'$ light is from the nucleus. HYPERZ uses Bruzual \& Charlot~(1993) spectral synthesis models and varies the redshift, age and extinction of the population. The elliptical galaxy model we use has an exponentially-declining star formation rate with an e-folding time of 1~Gyr; other models with longer star-formation timescales (S0--Sd and Im), as well as bursts, were also considered, but they do not significantly change the fitted redshift (essentially because the break between the V(F555W) and R points is assumed to be the redshifted 4000\AA~break).

For case (i), HYPERZ fits a 500~Myr old elliptical galaxy with $A_{\rm{V}}=2.40$~mag at $z=0.575$. For this redshift, we generated a grid of DUSTY models, with $\tau =10$, 20,..,100 (equivalent to $N_{\rm{H}} \simeq 10^{22}-10^{23}$\psqcm~for a Galactic dust:gas ratio, and appropriate for the moderately Compton-thin obscuration implied by the X-ray fluxes), $T_{\rm{in}}=100$, 200,..,1500\K, and $R=5$, 50, 250 and 1000. After integration over the {\em ISOCAM} filter bandpasses, the models were normalised to the observed 15\micron~flux density, and deemed acceptable if the 6.7/15 \micron~flux ratio fell within the range allowed by the errors on the {\em ISOCAM} data, if the 850\micron~flux fell below the {\em SCUBA} limit, and if the K$'$ flux was less than 20 per cent of that observed. Fig.~1 shows which models meet these criteria, and Fig.~2 the SED fits for three such cases, with $(R,\tau,T_{\rm{in}})=$(250,50,1500), (50,100,700) and (5,60,200). The implied optical-UV DUSTY input luminosity, 2--10\keV~X-ray absorption corrected luminosity (derived from the observed 2--7\keV~flux, assuming an intrinsic power-law spectrum with a photon index $\Gamma=2$), optical (2500\AA) to X-ray (2\keV) spectral index, $\alpha_{\rm{ox}}$, and the inner radius, $r_{\rm{in}}$, for each are shown in Table~1. The best fits to the portion of the SED covered by the {\em ISOCAM} data are obtained with hot dust (heated close to its sublimation temperature of 1500\K), with an inner radius within a parsec of the central engine; the implied $\alpha_{\rm{ox}}$ in this case is, however, somewhat flatter than the canonical value of 1.35 for quasars (Elvis et al.~1994). Figs.~1 and 2 also show, however, that models with warm dust at $T_{\rm{in}}=200$\K~cannot be ruled out at present; such models have inner radii of tens of \pc~and $\alpha_{\rm{ox}}$ values close to those of quasars. Discrimination between the warm and hot dust models would be possible using data at 70\micron; e.g. the 200\K~model shown in Fig.~2 has a 70\micron~flux density of 16~mJy, which is well within the capability of SIRTF (see e.g. Brandl et al.~2000).

For case (ii), HYPERZ fits a 3.5~Gyr old elliptical galaxy at z=0.505 with $A_{\rm{V}}=0.60$~mag. The need for DUSTY to reproduce at least half of the K$'$ light in this case means that the dust must be hot ($T_{\rm{in}}=1000-1500$\K) and not too optically thick at 2\micron~($\tau \sim 25-40$); $R$ is not very well constrained, with a value of 100 for $\tau=35$, although a more compact structure, $(R,\tau)=(5,40)$, is also acceptable. For $(R,\tau)=(100,35)$, we find $L_{\rm{2-10}}=4.6 \times 10^{43}$\ergps~and $\alpha_{\rm{ox}}=1.12$; the latter model SED is shown in Fig.~2.

We conclude that the best-fitting models are those in which the dust is hot, with an inner radius within $\sim 1$\pc~of the nucleus. Several models with warm dust (200\K) are also formally acceptable, but for $R=50$ some of these underpredict the {\em SCUBA} upper limit by only a small factor ($\sim 2$), which, if this source is representative of the new {\em Chandra} HXB sources, seems unlikely given the non-detection of large samples of them by {\em SCUBA} (see references in section~1). 

\begin{figure}
\psfig{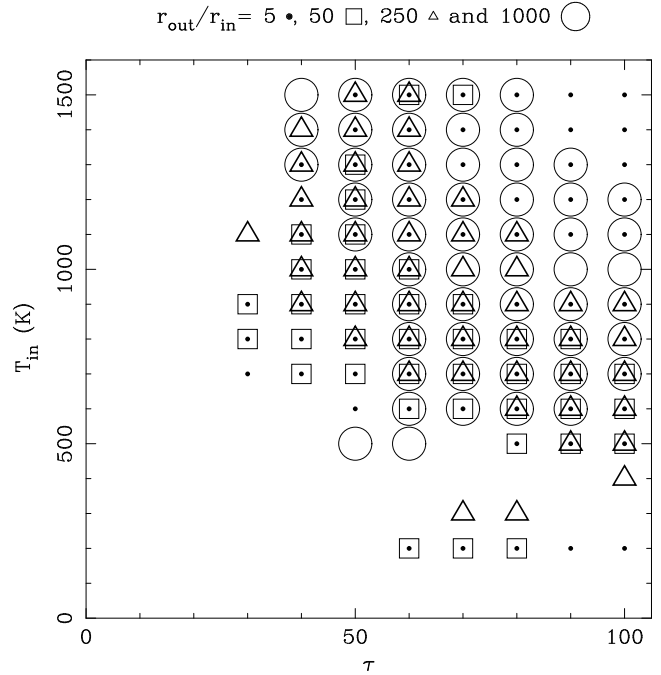}
\caption{\normalsize The points show which of the DUSTY grid models described in section 2.1 for source B case (i) are acceptable, for different values of $R=r_{\rm{out}}/r_{\rm{in}}$.}
\end{figure}

\begin{figure}
\psfig{figure=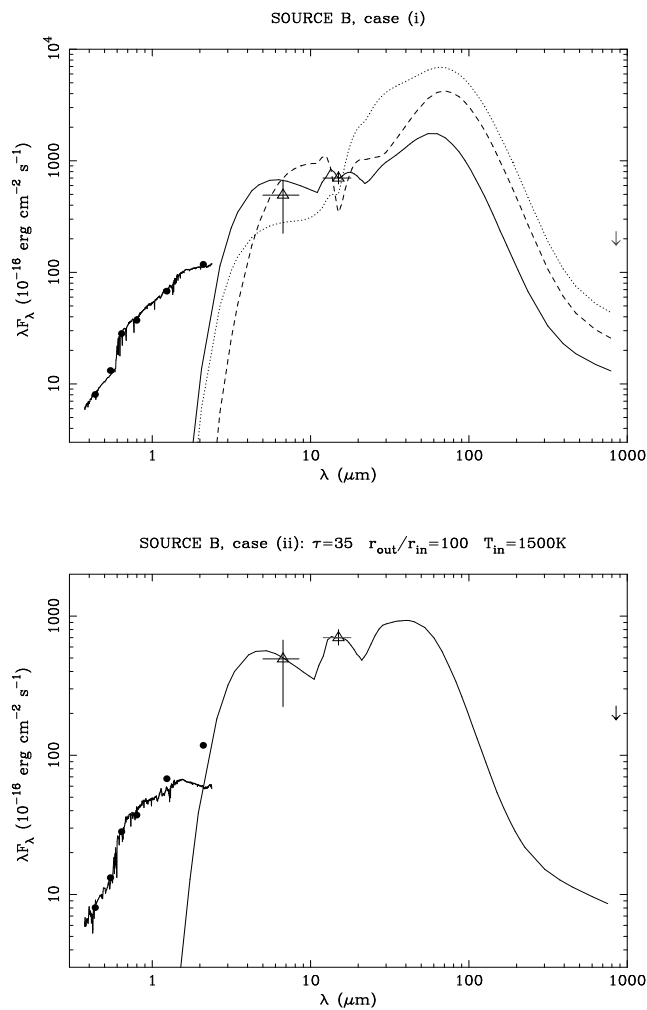,width=0.48\textwidth,angle=270}
\caption{\normalsize Observed and model SEDs for source B, for the cases where the galaxy produces essentially all (case i) and half (case ii) of the light at 2.1\micron. The circles and triangles are the optical and {\em ISOCAM} fluxes, respectively, from L\'{e}monon et al.~(1998); the arrow is the SCUBA upper limit from F00. The DUSTY models are normalised to the observed 15\micron~flux density; the optical points are overlaid with the best-fit HYPERZ model for the host galaxy. In the upper panel, the solid, dashed and dotted lines are the models with $(R,\tau,T_{\rm{in}})=$(250,50,1500), (50,100,700) and (5,60,200), respectively.}
\end{figure}

\begin{table*}
\caption{Details of three acceptable DUSTY models for source B case (i), spanning the range of allowed temperatures.}
\begin{tabular}{llll}
                                 		&  Model 1 &  Model 2 &  Model 3 \\  \\
$\tau$						&  50 	   &  100     &	  60 	\\  	 	
$T_{\rm{in}}$   (\K)				&  1500	   &  700     &   200   \\
$R=r_{\rm{out}}/r_{\rm{in}}$			&  250	   &  50      &   5     \\
$r_{\rm{in}}$ (\pc)				&  0.41    &  3.9     &   67     \\
$L(opt-UV)$ ($10^{44}$\ergps) $\dagger$		&  6.2	   &  11.3    &   19.2    \\
$L(2-10~\keV)$ ($10^{43}$\ergps) $\ddagger$	&  6.3	   &  7.5     &   6.5    \\
$\alpha_{\rm{ox}}$  $\star$				&  1.17    &  1.24    &   1.35   \\ \\
\end{tabular}

$\dagger$ Luminosity of the DUSTY input continuum. \\
$\ddagger$ Absorption-corrected 2--10\keV~luminosity. \\
$\star$ Implied optical to X-ray spectral index of the intrinsic continuum.
\end{table*}

\subsection{Source A (CXOUJ215334.0+174240)}
This is the brightest {\em Chandra} source in the A2390 field, coincident with a slightly reddened $L^{\star}$ mid-type spiral at a photometric redshift of $z=0.85 \pm 0.15$ (F00). From a total of $\sim 90$ counts, F00 performed crude X-ray spectral fitting and deduced an intrinsic $N_{\rm{H}}=(6 \pm 2)-(9 \pm 3) \times 10^{22}$\psqcm, with a de-absorbed $L_{\rm{2-10}} \simeq 2-3 \times 10^{44}$\ergps~(after correction for a lensing magnification by a factor of about 2), for $z=0.7-1$, respectively. It is thus an X-ray type II quasar.

It also has {\em ISOCAM} counterparts at 6.7 and 15\micron~(Altieri et al.~1999) but the source fluxes have not been published. We proceed on the assumption that they are equal to those of source B, which appears reasonable from the images in Altieri et al. 

For redshifts in the range 0.7--1.0, we find acceptable DUSTY models with $\tau=55-80$ (equivalent to the fitted $N_{\rm{H}}=6-9 \times 10^{22}$\psqcm~for a Galactic dust:gas ratio) for $R=100$ and $T_{\rm{in}}=1500$\K; they have $r_{\rm{in}}=0.55-1.0$\pc~and $\alpha_{\rm{ox}} \simeq 1.0$.

\section{DISCUSSION}
It is instructive to compare our results with the calculations of Granato, Danese \& Franceschini~(1997), who compared dust radiative transfer models with the infrared properties of Seyfert galaxies. They concluded that the moderately thick, extended tori of Granato \& Danese~(1994) (GD) (with $5 \leq A_{\rm{V}} \leq 80$~mag and outer radii of 10's to 100's of parsec), provided a better fit to the data than the thick, very compact models of Pier \& Krolik~(1992a) (PK) (with $A_{\rm{V}} \geq 800$~mag, all within a \pc). They also found that the observed X-ray absorption to the nucleus was much higher than that implied by the $A_{\rm{V}}$ obtained by modelling the infrared SED (for a Galactic dust:gas ratio), suggesting that much of the X-ray absorbing gas lies within the dust sublimation radius.

In terms of its optical depth and radial extent, the obscuring dust in these two Compton-thin, absorbed, X-ray background sources more closely resembles the model of GD than that of PK. Indeed, Granato, Danese \& Franceshini~(1997) predicted a significiant correlation between the HXB sources and those appearing in {\em ISO} surveys at 10--20\micron, precisely because the GD models are (at the most) only moderately optically thick over this wavelength range. An important difference, however, is that the GD tori have opening angles of 35--45 degrees, whereas the covering fraction of any torus-like structure in the HXB sources must be higher (e.g. Fabian \& Iwawasa~1999 deduce that 85 per cent of the accretion power must be absorbed). Note also that the DUSTY calculations assumed a spherical dust geometry. It is thus appropriate to examine how such a high space covering obscuration could be maintained against the dissipative forces which would tend to yield a flattened structure: at the sub-parsec scale inner radii found for sources A and B, Pier \& Krolik~(1992b) have shown that the pressure of the nuclear radiation on the grains can make the torus geometrically thick, by both reducing the vertical component of gravity, and if the torus is sufficiently clumpy, driving random motions of its constituent clouds. Fabian et al.~(1998) demonstrated that a space-covering obscuration could be maintained with energy input from a nuclear starburst within 100\pc, but the distinct lack of emission lines in many of the newly-discovered HXB sources appears to rule this out. More generally, the lack of AGN emission lines in such objects may be due to dust within the narrow line region clouds. Netzer \& Laor~(1993) showed that its presence could significantly suppress line emission, through absorption of the photoionizing continuum and destruction of line photons, and thereby account for the mismatch between the inferred covering factors of the broad and narrow line regions in classical AGN when dust is not considered (the broad line region lies just within the dust sublimation radius). Our findings suggest the presence of a high covering fraction of {\em dusty} gas in the new HXB sources.

Alternatively, the lack of emission lines and non-detections by {\em SCUBA} could be used to argue that the new HXB sources are not actually obscured AGN, but some other class of intrinsically hard source (e.g. ADAFs, as proposed for the HXB by Di Matteo \& Allen~1999); the present quality of the X-ray spectra is not high enough to discriminate between these two possibilities. Our present findings, however, strongly support the obscured AGN hypothesis, by demonstrating that the energy which is inferred to be absorbed in the optical--UV--X-ray range is reradiated in the infrared. In consequence, the results of deep {\em Chandra} surveys of well-studied {\em ISO} fields (e.g. from the ELAIS consortium; Rowan-Robinson et al.~1999), are eagerly awaited. Indeed, as noted by F00, 5 of the 6 {\em Chandra} sources discovered by Hornschemeier et al.~(2000) in the HDFN also have {\em ISOCAM} counterparts at 15\micron~listed by Aussel et al.~(1999) (CXOHDFN~123648.1+621309 is the only source not detected). 

\section*{ACKNOWLEDGMENTS} 
We are grateful to Z.~Ivezi\'{c}, M.~Nenkova and M.~Elitzur for making DUSTY publically available, and likewise to M.~Bolzonella, R.~Pell\'{o} and J.-M.~Miralles for HYPERZ. Carolin Crawford is thanked for helpful comments and assistance.

RJW thanks PPARC for support, and PG the Isaac Newton Trust and the Overseas Research Trust. ACF thanks the Royal Society for support.

{}


\begin{thebibliography}{}
\bibitem []{} Altieri B., et al., 1999, A\&A, 343, L65

\bibitem []{} Aussel. H., Cesarsky C.J., Elbaz D., Starck J.L., 1999, A\&A, 342, 313

\bibitem []{} Bolzonella M., Pell\'{o} R., Miralles J.-M., 2000, submitted to A\&A (astro-ph/0003380)

\bibitem []{} Brandl B., et al., 2000, to be published in the ASP Conf. Series ``From Darkness to Light'', eds. T. Montmerle, P. Andre, Cargese (astro-ph/0007300)

\bibitem []{} Brandt W.N., et al., 2000, AJ, 119, 2349

\bibitem []{} Bruzual G.A., Charlot S., 1993, ApJ, 405, 538

\bibitem []{} Crawford C.S., Fabian A.C., Gandhi P., Wilman R.J., Johnstone R.M., 2000, submitted to MNRAS (astro-ph/0005242)

\bibitem []{} Di Matteo T., Allen S.W., 1999, ApJ, 527, L21

\bibitem []{} Draine B.T., Lee H.M., 1984, ApJ, 285, 89

\bibitem []{} Elvis~M., et al., 1994, ApJS, 95, 1 

\bibitem []{} Fabian A.C., et al., 2000, MNRAS, 315, L8 (F00)

\bibitem []{} Fabian A.C., Barcons X., Almaini O., Iwasawa K., 1998, MNRAS, 297, L11

\bibitem []{} Fabian A.C., Iwasawa K., 1999, MNRAS, 303, L34

\bibitem []{} Fabian A.C., 1999, MNRAS, 308, L39

\bibitem []{} Fiore F., et al., 2000, to appear in New Astronomy (astro-ph/0003273)

\bibitem []{} Granato~G.L., Danese~L., 1994, MNRAS, 268, 235

\bibitem []{} Granato~G.L., Danese~L., Franceschini, 1996, ApJ, 460, L11

\bibitem []{} Granato~G.L., Danese~L., Franceschini, 1997, ApJ, 486, 147

\bibitem []{} Hornschemeier A.E., et al., 2000, submitted to ApJ (astro-ph/0004260)

\bibitem []{} Ivezi\'{c} Z., Nenkova M., Elitzur M., {\em User Manual for DUSTY}, 1999 (astro-ph/9910475)

\bibitem []{} L\'{e}monon L., Pierre M., Cesarky C.J., Elbaz D., Pell\'{o} R., Soucail G., Vigroux L., 1998, A\&A, 334, L21

\bibitem []{} Mushotzky R.F., Cowie L.L., Barger A.J., Arnaud K.A., 2000, Nature, 404, 459

\bibitem []{} Netzer H., Laor A., 1993, ApJ, 404, L51

\bibitem []{} Pier E.A., Krolik J.H., 1992a, (PK) ApJ, 401, 109

\bibitem []{} Pier E.A., Krolik J.H., 1992b, ApJ, 399, 23

\bibitem []{} Rowan-Robinson M., et al., 1999, ESASP, 427, 1011

\bibitem []{} Wilman R.J., Fabian A.C., 1999, MNRAS, 309, 862

\end{thebibliography}
\end{document}